\documentclass[aps,prl,superscriptaddress,draft]{revtex4}
\usepackage{amsmath}
\usepackage{amssymb}
\input{epsf}

\newcommand{\rem}[1]{}

\begin{document}

\title{A bird's eye view of quantum computers}
\author{Giuliano Benenti}
\email{giuliano.benenti@uninsubria.it}
\affiliation{CNISM, CNR-INFM \& Center for Nonlinear and Complex Systems, 
Universit\`a degli Studi dell'Insubria, Via Valleggio 11, 22100 Como, Italy}
\affiliation{Istituto Nazionale di Fisica Nucleare, Sezione di Milano,
via Celoria 16, 20133 Milano, Italy}
\author{Giuliano Strini}
\email{giuliano.strini@mi.infn.it}
\affiliation{Dipartimento di Fisica, Universit\`a degli Studi di Milano, 
Via Celoria 16, 20133 Milano, Italy}
\date{\today}

\begin{abstract}
Quantum computers are discussed in the general framework
of computation, the laws of physics and the foundations 
of quantum mechanics.
\end{abstract}
\maketitle

\section{Introduction}

Quantum computation \cite{qcbook,nielsen}
has gained widespread interest as a new 
interdisciplinary field of research, which benefits from the contributions
of physicists, computer scientists, mathematicians, chemists and 
engineers. A first reason for this interest is that the 
simulation of quantum many-body systems, a problem of utmost
importance in quantum chemistry and biochemistry, is a difficult 
task for a classical computer as the size of the Hilbert space 
grows exponentially with the number of particles. 
On the other hand, these quantum systems can be simulated by means 
of a quantum computer with resources that grow only polynomially 
with the system size (see \cite{lloyd} and \cite{lidar}). 
It is also remarkable that quantum mechanics
can help in the solution of basic problems of computer science,
such as the prime-factorization problem, for which quantum computation
provides an exponential speedup with respect to any known classical 
computation \cite{shor} (it is useful to note here 
that cryptographic schemes extensively used today such as RSA
are based on the conjecture that, given a composite odd positive
integer, it is ``hard'' to find its prime factors;
hence a large scale quantum computer, if constructed, would break 
the RSA encryption scheme). Of course, there are problems for which 
quantum computation leads to an algebraic speedup (for instance,
searching a marked item in an unstructured database \cite{grover}) 
or to no speedup at all. 

Besides efficiency of computations, 
an important question is whether the class of problems
that can be solved on a quantum computer is the same as on a classical
Turing machine. The answer is positive according to the standard
model of quantum computation, even though the possibility
of a different answer based 
on infinite-dimensional Hilbert spaces will be briefly mentioned later.

It is convenient to set the quantum computer within the general 
frame of the technological development of computer machines.
Indeed, miniaturization provides us with an intuitive way of
understanding why, in the near future, quantum mechanics will become
important for computation. The electronics industry
for computers grows hand-in-hand with the decrease in size of integrated
circuits. This miniaturization is necessary to increase
computational power, that is, the
number of floating-point operations per second (flops) a
computer can perform. 
Smaller size circuits boost computer power
because the communication between components is faster,
smaller active components are faster and at
the same time their density increases. 
In the 1950's, electronic computers based on vacuum-tube
technology were capable of performing approximately $10^3$ floating-point
operations per second, while nowadays there exist
supercomputers whose power is about $100$ teraflops ($10^{14}$ flops). 
Therefore, today's supercomputers are $10^{11}$ times more powerful
than $50$ years ago. It is then clear that it is more appropriate to
consider the growth rate of computer power rather than today's
computer power. The computational power has been growing at least
a factor of two every two years, for the last $50$ years.
This enormous growth of computational power
has been made possible owing to progress in
miniaturization, which may be quantified
empirically in Moore's law \cite{moore}. 
This law is the result of a remarkable
observation made by Gordon Moore in 1965: the number of
transistors on a single integrated-circuit chip doubles approximately
every $18-24$ months. This exponential growth has not yet
saturated and Moore's law is still valid. At the present
time the limit is approximately $10^8$ transistors per
chip and the typical size of circuit components is of the order
of $100$ nanometers. 
Extrapolating Moore's
law, one would estimate that within the next
10-20 years we will reach the atomic size for storing a single bit of
information. At that point, quantum effects will become
unavoidably dominant.
 
It is then natural to raise the following question: can one exploit
quantum mechanics to transmit and process information? 
The aim here is to build a quantum computer based on quantum logic, 
that is, it processes the information and performs logic operations 
in agreement with the laws of quantum mechanics. Quantum computers
were envisioned by Feynman in the 1982 (see \cite{hey}) and more
recently quantum information processing and communication 
has established itself as one of the hot topic fields in 
contemporary science. The great challenge is to build quantum
machines based on quantum logic, which process the information
and perform logic operations by exploiting the laws of quantum 
mechanics. The power of quantum computation is rooted in
typical quantum phenomena, such as the superposition of quantum states and
entanglement. There is an inherent quantum parallelism associated
with the superposition principle. In simple terms, a quantum computer 
can process a large number of classical inputs in a single run.
On the other hand, this implies a large number of possible outputs.
It is the task of quantum algorithms, which are based on
quantum logic, to exploit the inherent quantum parallelism of 
quantum mechanics to highlight the desired output. 
Therefore, besides the imposing technological challenges, 
to be useful quantum computers also require the development of
appropriate quantum software, that is, of efficient quantum
algorithms.

The purpose of these short notes is to give a flavor of the 
state of the art of quantum computation, focusing on open
problems and perspectives rarely discussed in the literature.
Only a few papers useful as entry points will be quoted. 
For a more complete bibliography, see the  
textbooks \cite{qcbook,nielsen}

\section{Quantum computers and quantum mechanics}

An operative quantum computer would allow us to perform 
countless experimental tests on the foundations of quantum
mechanics, at present only considered ``gedanke experimente''.
For instance, tests on the nonseparability of quantum mechanics
\cite{aspect} could be implemented between two 
entangled registers moving one with respect 
to the other. In order to test possible gravitational
effects on quantum states, one register
could be located on the earth's surface and the other 
in a satellite. 

The basic principles of quantum mechanics required for the working
of a quantum computer are very simple and can be conveniently 
summarized by the quantum circuit drawn in Fig.1.
Let us consider a $n$-qubit quantum computer, whose input state
is described by a density operator $\rho$ acting on the 
$n$-qubit Hilbert space of dimension $2^n$. The environment is
represented by $m$ additional qubits, initially in a (pure)
state $|0^m\rangle$. The overall system-environment evolution
is described by the unitary matrix $U$, which includes both the 
quantum gates and the undesired system-environment interaction
inducing decoherence. Finally, the environment
is disregarded (mathematically, we trace the overall state over
the environmental degrees of freedom) and $n^{\prime\prime}$ qubits 
are measured in the computational basis (say, the basis of
the eigenstates $|0\rangle$ and $|1\rangle$ of the Pauli 
spin operator $\sigma_z$). This measurement induces the wave function
collapse and we end up with the state $\rho^\prime$ for the remaining
$n^\prime=n-n^{\prime\prime}$ qubits. 

\begin{figure}
\centerline{\epsfxsize=9.cm\epsffile{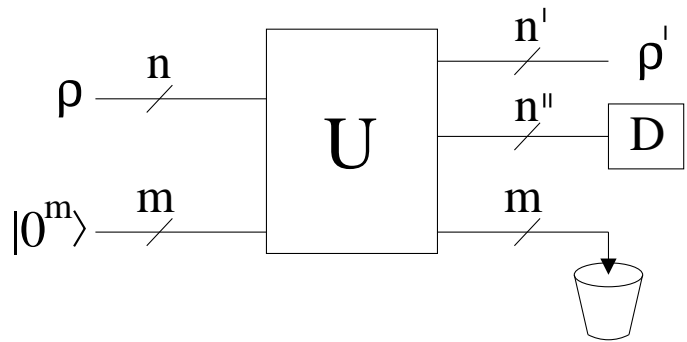}}
\caption{A quantum circuit summarizing the working of a quantum computer.}
\end{figure}

It should be stressed that the theoretical picture summarized in
Fig.1 only refers to quantum systems with finite-dimensional Hilbert spaces.
Furthermore, one should remember that there is an existing 
tension between quantum mechanics and the relativistic theory 
of gravitation, still not solved by quantum gravity
theories such as loop quantum gravity and superstrings theories.
Of course, one could assume quantum gravity effects to be
important only at length scales below the Planck scale 
(approximately $10^{-33}$ cm). On the other hand, given the
absence of a complete theory, it cannot be excluded that effects
would appear at a larger length scale, possibly hampering 
the construction of a quantum computer with many qubits. 
We emphasize that a quantum computer useful for the solution of
problems in quantum chemistry or for factoring integers beyond
the capabilities of a classical computer would require 
$10^2-10^4$ qubits. On the other hand, the predictions of 
quantum mechanics have been so far tested with up to $11$ 
significant digits (we refer here to the experiments measuring 
the magnetic moment of the electron). As the size of the Hilbert
space grows exponentially with the number of qubits, 
the implementation of a quantum computer with $10^2-10^4$ qubits 
would represent a verification of the validity of quantum 
mechanics much beyond the tests so far realized.

At any rate, even the first, few-qubit demonstrative experiments are remarkable,
not only for quantum computation but also for addressing
fundamental questions on quantum mechanics, such as the nature of the frontier
between quantum and classical worlds or the nature of quantum entanglement in
complex many-body systems. We can gain from quantum computation and 
information a better intuition of the weird world of quantum mechanics 
and of its most counterintuitive aspects. It is here useful to remind
the reader that even Schr{\"o}dinger wrote \cite{schrodinger} that
\emph{``We never experiment with just one electron or atom or (small)
molecule. In thought-experiments we sometimes assume that we do; this
invariably entails ridiculous consequences \dots\ we are not experimenting with
single particles, any more than we can raise Ichthyosauria in the zoo''}.
It is absolutely remarkable that only fifty years
later experiments on single electrons, atoms and molecules are
routinely performed in laboratories all over the world, for instance
in few-qubit implementations of quantum algorithms.

\section{The problem of decoherence}

In practice, any quantum system is open; namely, it is
never perfectly isolated from the environment. The word decoherence, used
in its broader meaning, denotes any quantum noise process
due to the unavoidable coupling of the system to the environment.
Decoherence can be considered as the ultimate threat to the actual
implementation of any quantum computation or communication protocol.
Indeed, decoherence invalidates the quantum superposition principle, which
lies at the heart of the potential power of any quantum algorithm.
On the other hand, any quantum information protocol must end up
with a measurement, converting quantum states into classical outcomes, 
and decoherence plays a key role in this quantum measurement process.
The decoherence problem is very complex, 
both theoretically and experimentally.
To give an idea of its complexity it is sufficient to remind
the reader that in the most general evolution of a $n$-qubit 
quantum computer (see Fig.1) the action of the environment is 
described by $n^4-n^2$ real parameters; that is, $12$ parameters
for a single qubit, $240$ for two-qubits, and so on. The main 
problem is that it appears difficult to attach a concrete physical
meaning to all these parameters for more than one qubit. 
In the simplest case, $n=1$, the $12$ parameters are associated
to rotations, deformations and displacements of the 
single-qubit Bloch sphere. On the other hand, even in this simple
case quantum noise is usually described by means of phenomenological
theories involving a smaller number of parameters, for instance
just the dephasing and relaxation time scales. 
It is indeed very difficult, even for a single qubit, to perform
experimentally a complete ``tomography'' of noise and so 
determine all the above $12$ parameters.

According to the standard theory of quantum mechanics decoherence is 
a technical, though very difficult problem. In other words, 
there are no fundamental limitations to the generation and 
persistence of arbitrarily complex entangled states. However,
it is worth mentioning that there exist interpretations of 
quantum mechanics in which ``intrinsic decoherence'' for isolated
systemes of particles is introduced, see for instance \cite{ghirardi}.
In general, the problem of decoherence is strictly connected 
to the emergence of classicality in a world
governed by the laws of quantum mechanics. 
Even though this latter problem
has fascinated scientists since the dawn of quantum mechanics,
there exists as yet no theory solving it in a 
fully satisfactory way.

\section{Quantum error correction}

If errors are not corrected, the error per gate that can be 
tolerated tends to zero when the number of gates tends to
infinity. It is then clear that, to allow arbitrarily long 
computations with reasonable physical resources, error
correction is necessary.
The question is how to protect quantum information 
from errors. It is indeed evident that coherent superposition
of quantum states are very fragile and prone to decoherence.
However, quantum error-correcting codes fighting the effects 
of noise have been developed. These codes are very efficient 
for the correction of single-qubit errors, acting independently
on each qubit. On the other hand, when several qubits couple
identically to the environment (collective decoherence), it is 
possible to encode the information in decoherence-free subspaces.
However, it is clear from the above considerations on decoherence
that the problem is in general complex. For instance, when dealing
with two-qubit errors one has to distinguish between two
single-qubit errors (described by $2\times 12$ parameters) and
``true'' two-qubit errors, described by $240-2\times 12$ parameters.
This latter class of errors has not received enough attention
in the literature, so far.

\section{Experimental implementations}

The great challenge of quantum computation is to experimentally
realize a quantum computer. 
We need approximately $30-100$ qubits and thousands of quantum gates 
to build a quantum simulator capable of solving quantum mechanical problems
beyond the capabilities of present-day supercomputers. On the other hand,
thousand of qubits are required to outperform existing classical
computers in computational problems such as molecular structure
determination. Many requirements must be fulfilled
in order to achieve this imposing objective.
We require a collection of two-level quantum systems that can be
prepared, manipulated and measured at will. That is, our purpose
is to be able to control and measure the state of
a many-qubit quantum system. A useful quantum computer must be
scalable since we need a rather large number of qubits to perform
non-trivial computations. In other words, we need the quantum analogue
of the integrated circuits of a classical computer.
Qubits must interact in a controlled way if we
wish to be able to implement a universal set of quantum gates.
Furthermore, we must be able to control the evolution of a large number of
qubits for the time necessary to perform many quantum gates.

Given the generality of the requirements to build a
quantum computer, many physical systems might be good candidates,
and very interesting few-qubit experiments have been performed
with nuclear magnetic resonance techniques applied to nuclear
spins of molecules in the liquid state, with neutral atoms and
photons interacting in a resonant cavity, with cold ions in a trap
and with solid-state qubits.
Qubits made out of solid-state devices may offer great advantages
since fabrication by established lithographic methods
allows for scalability (at least in principle). Moreover, another
important feature of solid-state devices is their flexibility in
design and manipulation schemes. Indeed, in contrast to
``natural'' atoms, ``artificial'' solid-state atoms can be
lithographically designed to have specific characteristics such as a
particular transition frequency. This tunability is an important advantage over
natural atoms. Finally, solid-state qubits are easily embedded in
electronic circuits and can take advantage of the rapid
technological progress in solid-state devices as well as of 
continuous progress in the field of nanostructures. On the other hand, 
it should be remarked that there is a great variety of decoherence 
mechanisms, still not well understood, in solid-state devices.
When the problem of decoherence is taken into account for a complex many-qubit
system, which we require to perform coherent controlled evolution, 
then large-scale quantum
computers appear unrealistic with present technology. On the other hand, we
should bear in mind that also the technological development of the classical 
computer took decades and that breakthroughs (such
as the transistor was for the classical computer) are always
possible.

Quantum technologies are not limited to quantum computation
but include other relevant applications such as quantum cryptography.
It appears probable that in the near future quantum
cryptography will be the first quantum-information protocol to find
commercial applications, and indeed quantum cryptosystems are
already sold. Here the question is how extensive the
market will be and this will largely depend on the transmission rates, at
present limited to the kHz range. The development of fast
single-photon sources and high-efficiency detectors is required to
improve significantly the transmission rates, thus broadening the prospects of
quantum cryptography.

\section{Are quantum computers the ultimate frontier?}

It is worth mentioning that some years ago computers  
using electronic and optical processes between molecules
were envisaged (see \cite{hameroff} and \cite{carter}) and 
presented as ultimate computers
(note that these ideas have also stimulated a very interesting experimental 
activity). However, it should be stressed that 
molecular computers are based on
classical Boolean logic. In contrast, quantum computers
replace the laws of classical physics applied to computation with the more
fundamental laws of quantum mechanics.
Computers are physical devices, whose working 
is governed by physical laws, so 
the question whether quantum computers are the ultimate frontier
is strictly related to the question whether quantum mechanics
is the theory describing all physical phenomena at all scales.

If we remain in the framework of the quantum systems with 
finite-dimensional Hilbert spaces, it is possible to simulate a quantum
computer by means of a classical one. Therefore, the class of 
all functions computable by a quantum computer is equivalent to the 
class of all functions computable by a classical Turing machine. 
Of course, there might be for certain problems an exponential gain 
in time and/or memory resources. This means that a large-scale 
quantum computer could solve problems in practice beyond the reach
of classical Turing machines or, in general, of classical computers 
based on classical physics. Indeed, it is clearly impossible 
to implement a Turing machine with a tape divided into 
$10^{80}$ cells, this number being comparable to the number of 
nucleons in our universe. On the other hand, a quantum register
of $300$ qubits would overcome this limit.

At present, there are no experimental results in contradiction with 
the postulates of quantum mechanics. However, we cannot exclude that
in the future new experiments could require modifications of the 
foundations of quantum mechanics, perhaps supporting more
powerful computers. It is therefore worth mentioning some
results beyond ``standard'' quantum computation:
\begin{itemize}
\item
If the superposition principles of quantum mechanics might
be violated, that is, if the time evolution of quantum states
might be nonlinear, then quantum computers could be used to
solve \textbf{NP}-complete problems in polynomial time
\cite{abrams}.
The computational class \textbf{NP} is the class of problems
whose solution can be verified in polynomial time.
A problem in the class 
\textbf{NP} is \textbf{NP}-complete if any 
problem in \textbf{NP} is reducible to it by means of a mapping
involving only polynomial resources. We stress that no known polynomial 
time algorithms for the solution of \textbf{NP}-complete problems
exist, both on a classical and on a quantum computer based on
standard, linear quantum mechanics. 
\item
Kieu's algorithm \cite{kieu}. 
It has been claimed that, using infinite-dimensional Hilbert spaces
with special spectral properties of the Hamiltonian operator, 
it is possible to solve the Hilbert's tenth problem \cite{hilbert}.
Note that this problem is equivalent to the Turing halting problem
and therefore noncomputable on a classical as well as on a 
standard quantum computer.
\end{itemize}

\section{Final remarks}

In order to evaluate the future impact of quantum 
computation, the main question under discussion is:
is it possible to build a useful
quantum computer that could outperform existing classical
computers in important computational tasks? And, if so, when?
The difficulties are huge.
Besides the problem of decoherence, we should also remark on
the difficulty of finding new and efficient quantum algorithms. 
We know that problems such as molecular structure determination 
can be solved efficiently on a
quantum computer, but we do not know the answer to the following fundamental
question: What class of problems could be simulated efficiently on a quantum
computer? 
Quantum computers open up fascinating prospects, but it does
not seem likely that they will become a reality with practical
applications in a few years. How long might it take to develop the
required technology? Even though unexpected technological breakthroughs
are, in principle, always possible, one
should remember the enormous effort that was necessary in order to develop
the technology of classical computers.

We can certainly say that the computational power at our disposal 
increased enormously over the years and the computer hardware changed
from pebbles to mechanical and
electromechanical computers, vacuum tubes,
up to transistors and integrated circuits. 
The more optimistic predictions always underestimated the 
development of computers and we might hope the same will happen
for quantum computers. 
Of course, the time when a quantum computer will be on the desk 
in our office is uncertain. In any event, what is certain is that 
we are witnessing the emergence of a 
very promising field of investigation in physics, mathematics
and computer science.


\begin{thebibliography}{99}

\bibitem{qcbook} G. Benenti, G. Casati and G. Strini,
\textit{Principles of quantum computation and information},
Vol. I: Basic concepts (World Scientific, Singapore, 2004);
Vol. II: Basic tools and special topics 
(World Scientific, Singapore, 2007).

\bibitem{nielsen} M.A. Nielsen and I.L. Chuang,
\textit{Quantum computation and quantum information}
(Cambridge University Press, Cambridge, 2000).

\bibitem{lloyd}
S. Lloyd, \textit{Universal quantum simulators},
Science \textbf{273}, 1073 (1996).

\bibitem{lidar}
D.A. Lidar and H. Wang, \textit{Calculating the thermal rate 
constant with exponential speedup on a quantum computer},
Phys. Rev. E \textbf{59}, 2429 (1999).

\bibitem{shor}
P Shor, \textit{Polynomial-time algorithms for prime factorization 
and discrete logarithms on a quantum computer},
SIAM J. Sci. Statist. Comput. \textbf{26}, 1484 (1997).

\bibitem{grover}
L.K. Grover, \textit{Quantum Mechanics helps in searching 
for a needle in a haystack},
Phys. Rev. Lett. \textbf{79}, 325 (1997).

\bibitem{moore}
G.E. Moore, \textit{Cramming more components onto integrated circuits},
Electronics, \textbf{38}, N. 8, April 19 (1965).

\bibitem{hey}
A.J.G. Hey (Ed.),
\textit{Feynman and computation}
(Westview Press, 2002).

\bibitem{aspect}
A. Aspect, \textit{Proposed experiment to test the nonseparability
of quantum mechanics},
Phys. Rev. D \textbf{14}, 1944 (1976).

\bibitem{schrodinger}
E. Schr{\"o}dinger, \textit{Are there quantum jumps? Part~II},
Brit. J. Phil. Sci., \textbf{3}, 233 (1952).

\bibitem{ghirardi}
G.C. Ghirardi, A. Rimini, and T. Weber, \textit{Unified dynamics 
for microscopic and macroscopic systems}, 
Phys. Rev. D \textbf{34}, 470 (1986).

\bibitem{hameroff}
S.R. Hameroff, \textit{Ultimate computing}
(North-Holland, 1987).

\bibitem{carter}
F.L. Carter, \textit{Molecular electronic devices},
(Marcel Dekker, Inc., 1982).

\bibitem{abrams}
D.S. Abrams and S. Lloyd, \textit{Nonlinear quantum mechanics
implies polynomial-time solution for \textbf{NP}-complete
and \textbf{\#}\textbf{P} problems},
Phys. Rev. Lett. \textbf{81}, 3992 (1998). 

\bibitem{kieu}
T.D. Kieu, \textit{Computing the noncomputable}, 
Contemporary Physics \textbf{44}, 51 (2003).

\bibitem{hilbert}
D. Hilbert,
\textit{Mathematische Probleme},
G\"ottingen Nachrichten, 1900, pp. 253 - 297.
English translation (by M.W. Newson):
\textit{Mathematical Problems}: Lecture delivered before the International
Congress of Mathematicians at Paris in 1900,
Bulletin of the American Mathematical Society \textbf{8}, 437 
(1902).
Reprinted in:
\textit{Mathematical Developments Arising from Hilbert Problems},
Felix Brouder (Ed.), American Mathematical Society, 1976.

\end{thebibliography}
\end{document}